\documentclass{article}

\usepackage[english]{babel}

\usepackage[letterpaper,top=2cm,bottom=2cm,left=3cm,right=3cm,marginparwidth=1.75cm]{geometry}

\usepackage{amsmath}
\usepackage{float}
\usepackage{graphicx}
\usepackage{cite}
\usepackage[colorlinks=true, allcolors=blue]{hyperref}
\usepackage{subfig}
\usepackage{authblk}

\providecommand{\keywords}[1]
{
  \small	
  \textbf{\textit{Keywords---}} #1
}

\title{Convolutional Neural Network-Based Image Watermarking using Discrete Wavelet Transform}
\author[1$\dagger$]{Alireza Tavakoli}
\author[2$\dagger$]{Zahra Honjani}
\author[2*]{Hedieh Sajedi}
\affil[*]{Correspondence: \href{mailto:hhsajedi@ut.ac.ir}{hhsajedi@ut.ac.ir}}
\affil[1]{Department of Computer Engineering, School of Electrical and Computer Engineering, College of Engineering, University of Tehran, Tehran}
\affil[2]{Department of Computer Science, School of Mathematics, Statistics and
Computer Science, College of Science, University of Tehran, Tehran}
\affil[$\dagger$]{These authors contributed equally to this work.}
\date{}                     
\begin{document}
\maketitle

\begin{abstract}
With the growing popularity of the Internet, digital images are used and transferred more frequently. Although this phenomenon facilitates easy access to information, it also creates security concerns and violates intellectual property rights by allowing illegal use, copying, and digital content theft. Using watermarks in digital images is one of the most common ways to maintain security. Watermarking is proving and declaring ownership of an image by adding a digital watermark to the original image. Watermarks can be either text or an image placed overtly or covertly in an image and are expected to be challenging to remove. This paper proposes a combination of convolutional neural networks (CNNs) and wavelet transforms to obtain a watermarking network for embedding and extracting watermarks. The network is independent of the host image resolution, can accept all kinds of watermarks, and has only 11 layers while keeping performance. Performance is measured by two terms; the similarity between the extracted watermark and the original one and the similarity between the host image and the watermarked one.
\end{abstract}

\keywords{watermarking, convolutional neural networks, wavelet transform, neural networks}

\section{Introduction}

Protecting art ownership was always an issue since painters used to sign their works, which was not safe. Due to widespread Internet use, artistic works are becoming digital, so the threat of privacy, copyright, and illegal usage has increased. Authentication is not limited to artistic works but includes identity document (ID) cards and source tracking. In order to solve these issues, digital watermarking is an effective method of hiding information that can be used for authentication purposes.

The digital watermarking technique involves embedding information (watermark) into the content and extracting it when needed. Various attacks could be applied to the watermarked content to damage or remove the embedded watermark, which shows the importance of robust watermarking. Even though watermarking can be applied to a wide range of content, including video \cite{Patel2021}, audio \cite{Joshi2018}, or images, in this paper, we will focus on images.

Until recently, digital watermarking was based on deterministic algorithms. Typical methods are based on embedding the watermark in one of the discrete cosine transform (DCT) \cite{Xu2011}, discrete wavelet transform (DWT) \cite{two, Pradhan2020, Giri2018}, discrete Fourier transform (DFT) \cite{three}, or quantization index modulation \cite{four}, or a combination of these \cite{sajediwater, one}. 

Due to the inflexibility of deterministic algorithms, small changes in watermarked images lead to inaccurate extracted watermarks. The inflexibility leads to using neural networks as a robust technique for digital watermarking \cite{five, six}. Even though these methods solve the robustness problem, some non-deterministic networks still perform poorly, and others have complex structures. 

This paper investigates the combination of deterministic and neural network methods to improve accuracy and performance. As the results suggest, we have improvements in some attacks. In addition, the proposed network enhances the quality of watermarked images by using DWT as a preprocessing technique.

The rest of the paper is organized as follows: In Section \ref{sec:relate_works}, we discuss relevant previous studies. Then in Section \ref{sec:discrete_wavelet_transform}, we outline the basic concept, wavelet transform, used for the proposed algorithm. The proposed network structure is explained in Section \ref{sec:proposed_method}. Section \ref{sec:experiments} discusses the training technique, the experimental results, and the comparison with literature \cite{six}, and \cite{seven}. Moreover, this paper is concluded in Section \ref{sec:conclusion}.

\section{Related Works \label{sec:relate_works}}
Various image watermarking schemes have previously been done to provide robustness, authenticity, and imperceptibility. Earlier, we mentioned some papers proposing watermarking algorithms which use DWT, DFT, and neural networks. Here we will explore some of them in more detail. 

In paper \cite{two}, original images and watermarks are split into RGB channels to be used as inputs in the other embedding process. In the original image, the blue channel has been subjected to DWT and Singular Value Decomposition (SVD). A first Arnold transformation is applied to the blue channel of the watermark image, which acts as a key. In addition to Arnold's transformation, SVD is applied to the watermark image, and then the image is embedded with the watermark. In addition to being semi-blind, this method is also robust against attacks due to the use of the key during the watermark extraction. 

As another deterministic approach, Xiangui Kang proposed a blind DWT-DFT composite image watermarking algorithm that is robust against both affine transformation and Joint Photographic Experts Group (JPEG) compression \cite{one}. This paper used watermark structure, two-dimensional interleaving, and synchronization techniques to improve the robustness.

In recent years, residual blocks, fully connected layers, DCT layers, and CNN have been used as watermarking networks. An end-to-end autoencoder in ReDMark \cite{six} simulates a DCT layer in its deep network. The concatenation of the input image and watermark is used as input to the DCT layer, and after transformation, the embedding is performed. The proposed structure is robust against JPEG attacks since the DCT layer is used.

Jae-Eun Lee proposed a robust watermarking neural network that included an attack simulation and was adjusted to the resolution of the host image and the watermark \cite{seven}. It is composed of convolutional layers with average pooling layers. This method also includes a strength factor to adjust the tradeoff between invisibility and robustness. This method uses the images' luma component (Y) to hide the watermark. A watermark is converted into a layer the same size as the original image. The watermark is then merged with the processed image in the leading network. Despite the impressive results, the proposed network is relatively complex. In contrast to mentioned articles, this paper presented a non-deterministic method with higher robustness than discussed papers.

\section{Discrete Wavelet Transform \label{sec:discrete_wavelet_transform}}
\begin{figure}[H]
    \centering
    \subfloat[\centering Scaling function]{{\includegraphics[width=6cm]{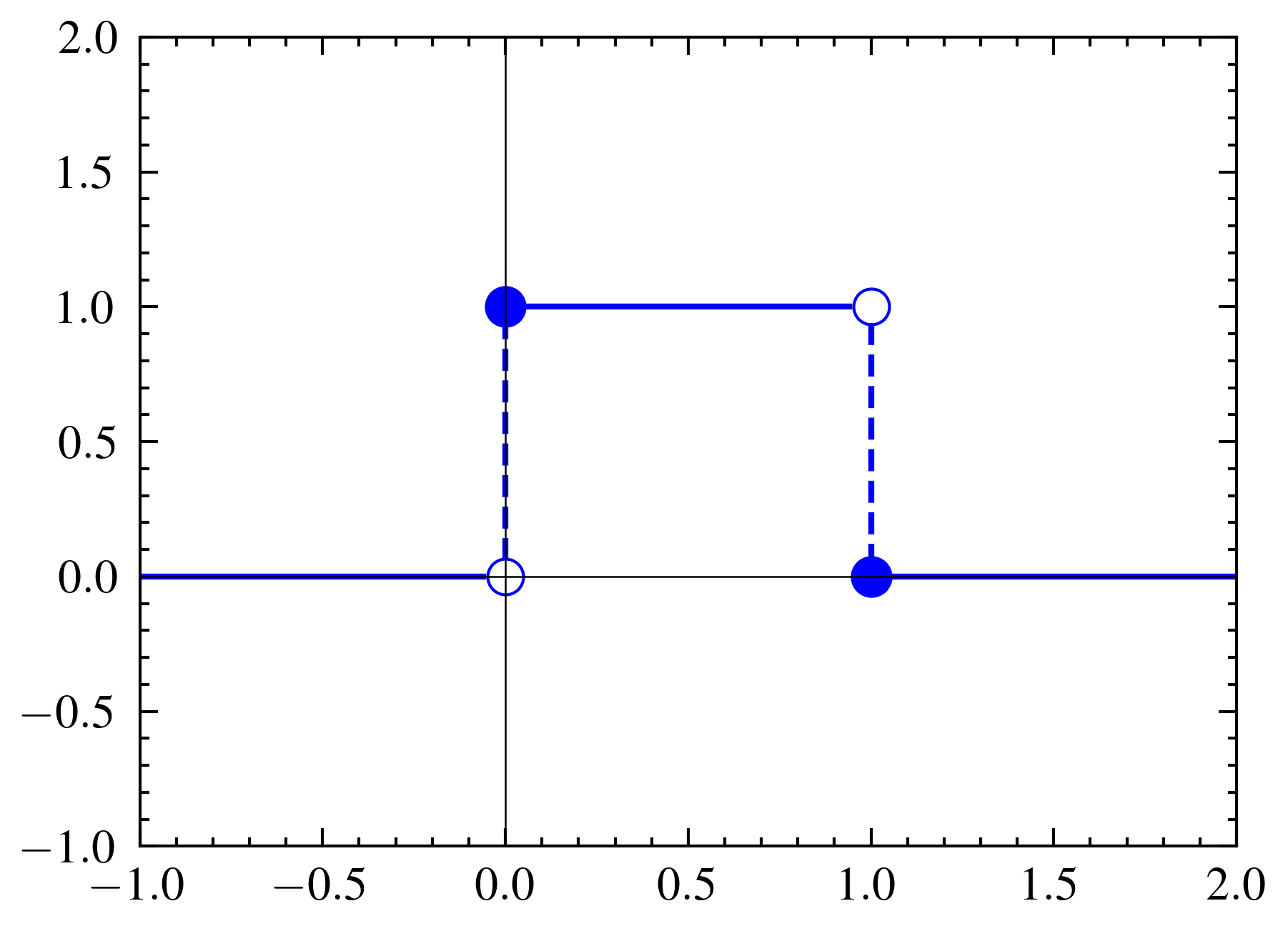} }}%
    \qquad
    \subfloat[\centering Mother function]{{\includegraphics[width=6cm]{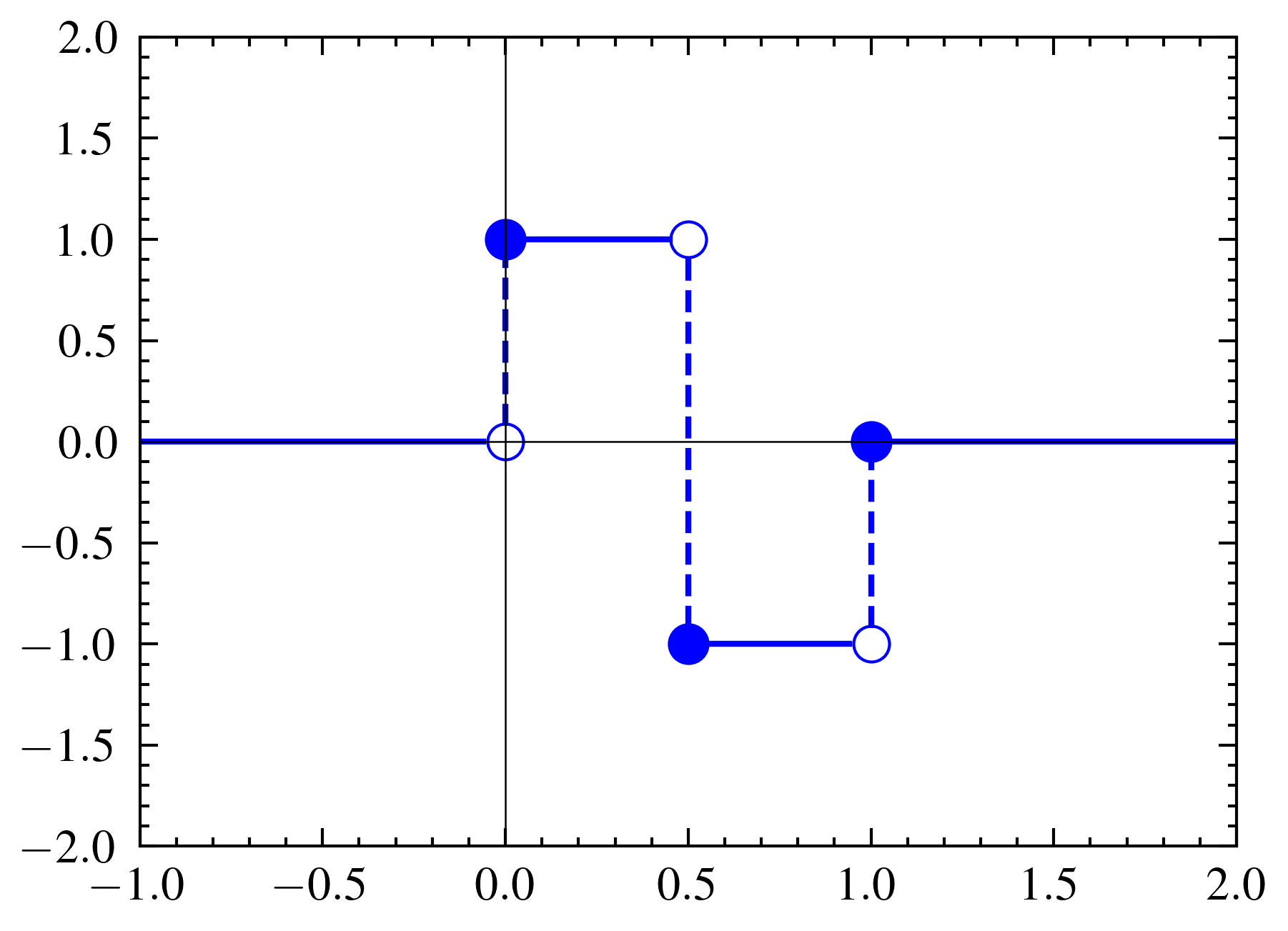} }}%
    \caption{Haar's functions.}%
    \label{fig:example}%
\end{figure}

Each image has smooth regions interrupted by edges. Edges provide much information about the image. In the Fourier transform, a signal is decomposed into sin waves that are not localized in time. Therefore, Fourier transforms cannot efficiently represent sudden changes (particularly edges), and wavelet transforms are preferred. 
A wavelet transform is a way of representing a function by a specific orthonormal series. A wavelet is a wave-like oscillation that is localized in time. Different types of wavelet transform exist based on the mother function $\psi(t)$ and the scaling function $\varphi(t)$. One of the most widely-used wavelets is Haar whose $\psi(t)$ and $\varphi(t)$ are shown in Figure \ref{fig:example}.

In addition, the discrete wavelet transform decomposes an image into four wavelet subbands, LL, LH, HL, and HH, which were examined independently to compare their capacity for hiding information. The results are demonstrated in Section \ref{sec:experiments} \cite{waveref}.

\section{Proposed Method \label{sec:proposed_method}}

\begin{figure}[H]
    \centering
    \includegraphics[height=6cm]{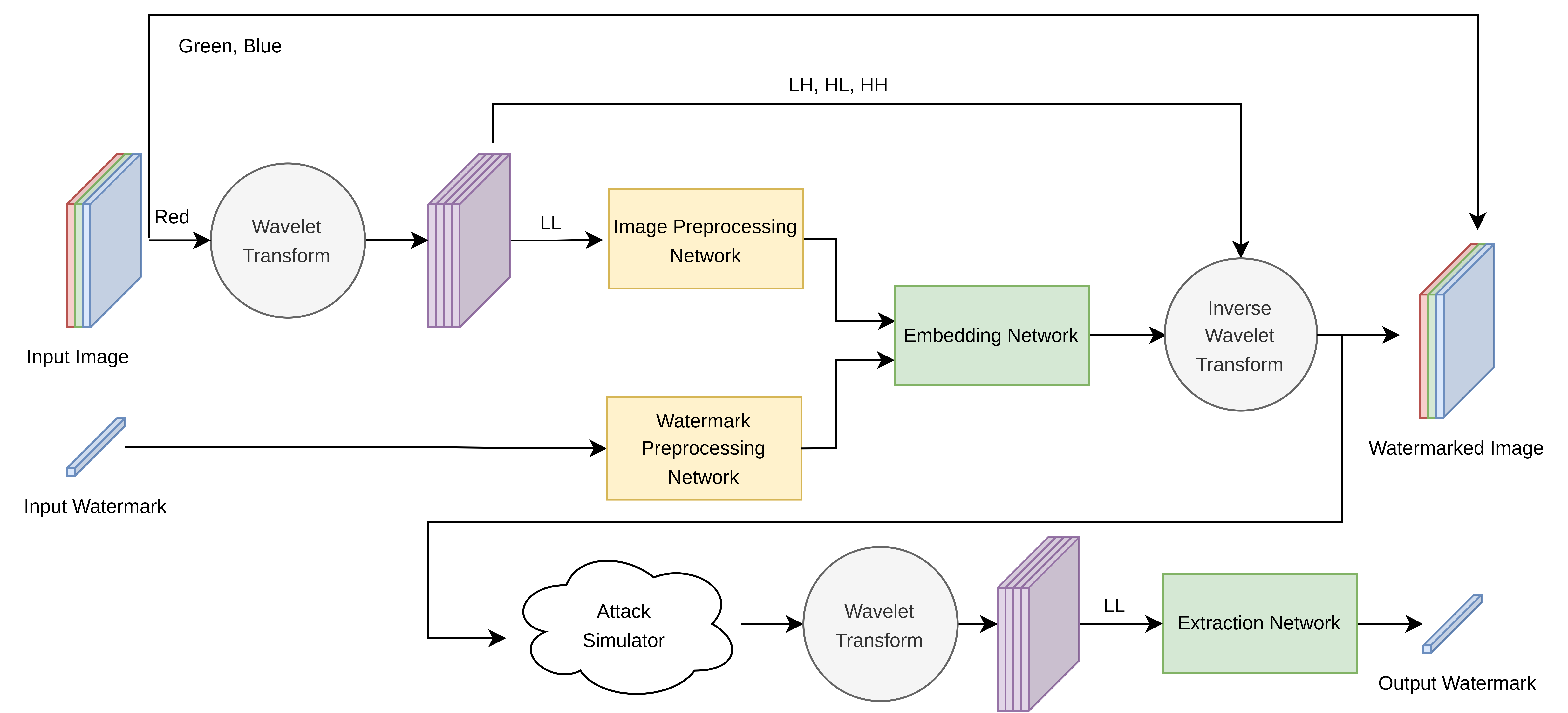}
    \caption{Network structure.}%
    \label{fig:overal_network}%
\end{figure}

The proposed network consists of 5 main parts; watermark preprocess network, image preprocess network, embedding network, attack layer, and extraction network. Throughout this paper, we will discuss each part in detail, but Figure \ref{fig:overal_network} shows the overall structure of the network. In addition to these parts, we also use the wavelet transform for embedding and extraction. Different shapes of images can be accepted through this structure because it is based on CNN layers. The codes of the proposed method are available on GitHub\footnote{\href{https://github.com/alirezatwk/Convolutional-Neural-Network-Based-Image-Watermarking-using-Discrete-Wavelet-Transform}{https://github.com/alirezatwk/Convolutional-Neural-Network-Based-Image-Watermarking-using-Discrete-Wavelet-Transform}}. Since DWT must be implemented in the model, we have used \cite{wavetf} to implement DWT layers that can run on the same hardware as the rest of the machine learning pipeline without impacting performance.

\subsection{Host Image Preprocessing Network}
\begin{figure}[H]
    \centering
    \includegraphics[height=4.5cm]{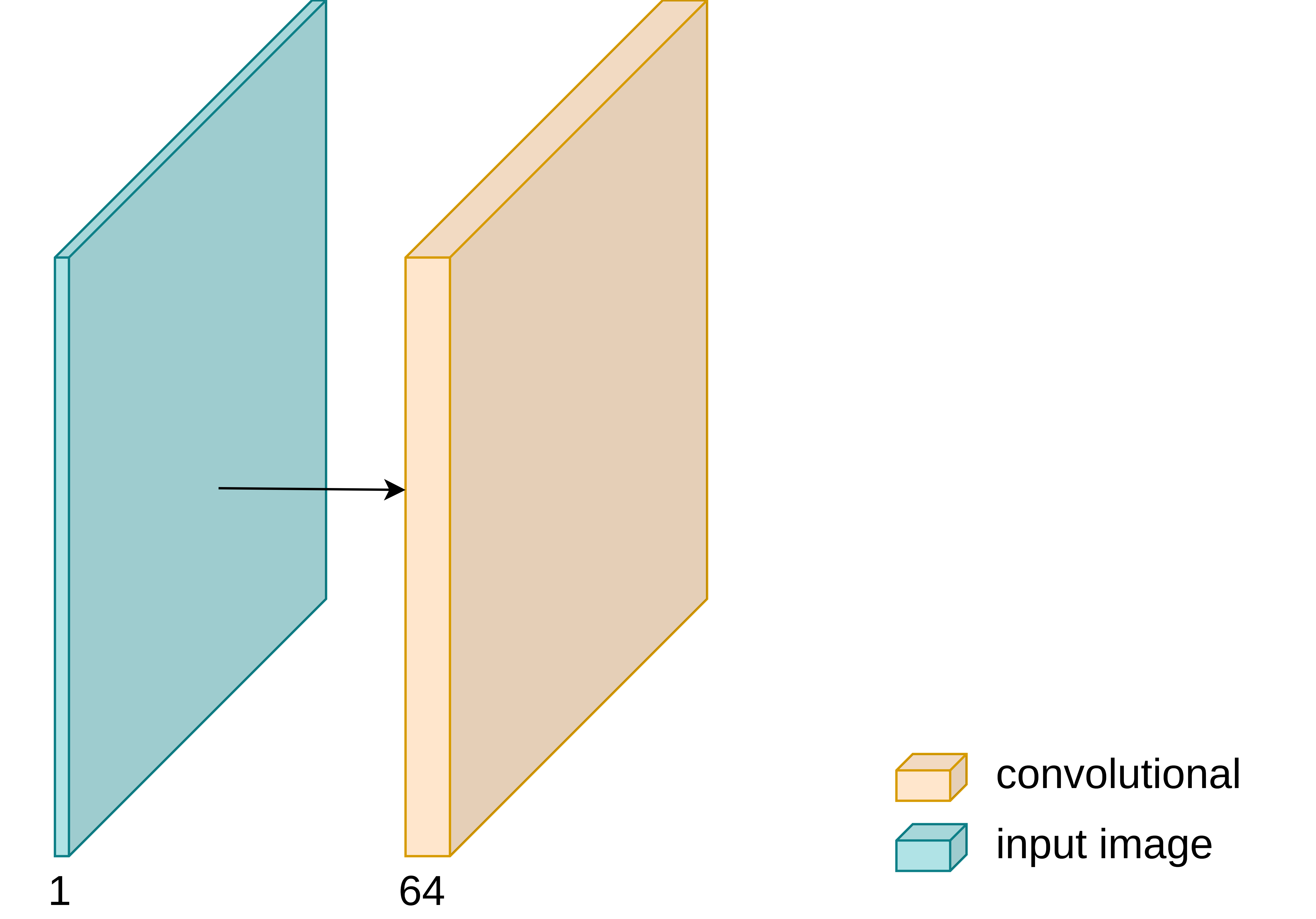}
    \caption{Host image preprocessing network structure.}%
    \label{fig:pre_image_network}%
\end{figure}

We select the first channel of the original image and scale it into the $[0,1]$ interval. Next, it is decomposed by the Haar wavelet. Then the LL layer is used as the input of the host image preprocessing network. The LL layer is fed into a convolutional layer with 3x3 filters whose strides equal one. According to Figure \ref{fig:pre_image_network}, the layer consists of 64 filters with the same padding so that the output shape is the same as the host image.

\subsection{Watermark Preprocessing Network} 
\begin{figure}[H]
    \centering
    \includegraphics[height=4.5cm]{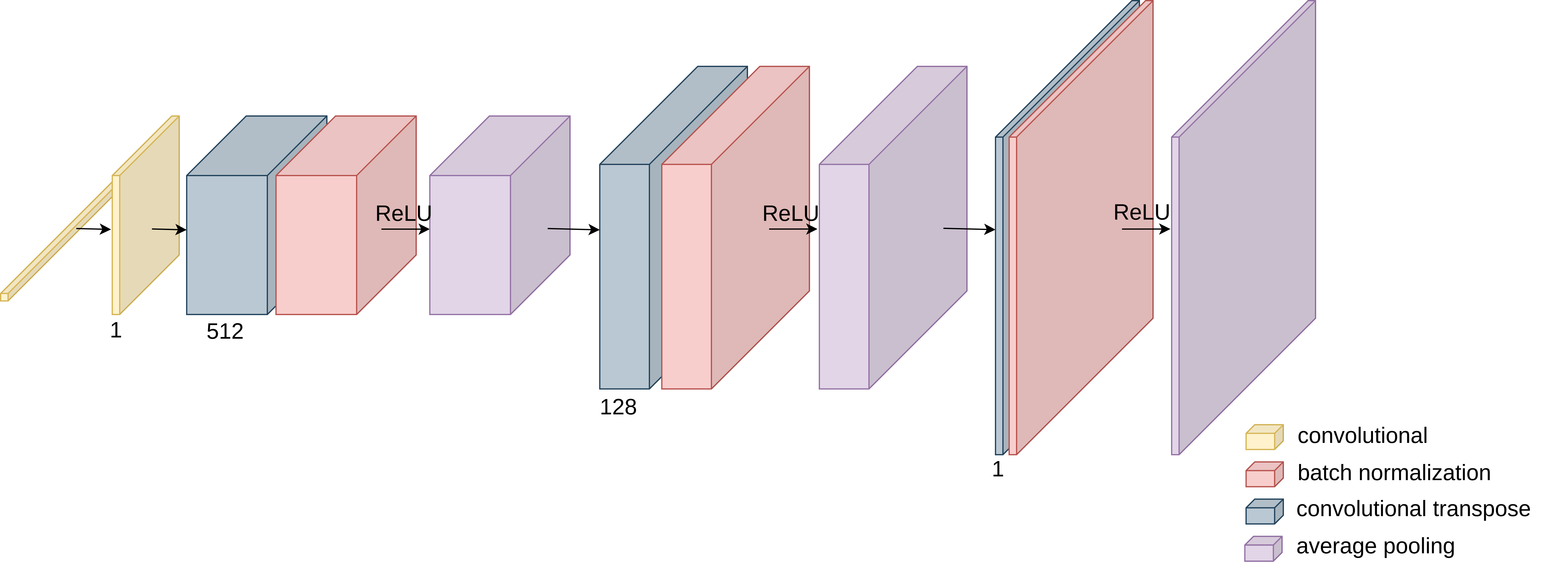}
    \caption{Watermark preprocessing network structure.}%
    \label{fig:pre_watermark_network}%
\end{figure}

As shown in Figure \ref{fig:pre_watermark_network}, the first step is to reshape array-like watermarks to 2D watermarks so that applying convolutional layers will be possible. The preprocessing network includes three network blocks. The blocks comprise convolutional transpose, batch normalization, ReLU activation, and average pooling. Convolutional transposes have 512, 128, and 1 filters, respectively. The output’s data type and shape are the same as the host image, so we can concatenate them in the following networks.

\subsection{Embedding Network}
\begin{figure}[H]
    \centering
    \includegraphics[height=4.5cm]{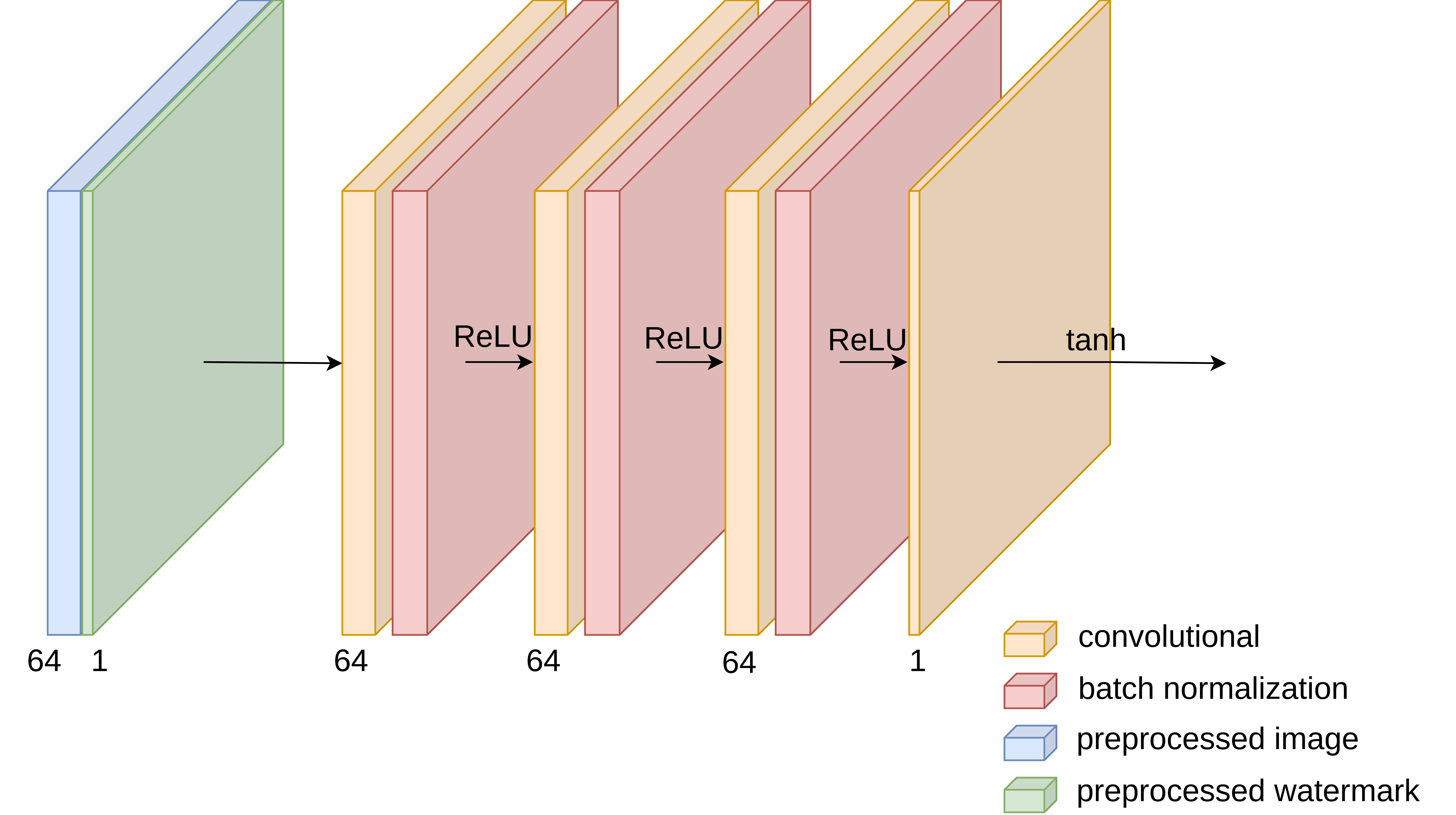}
    \caption{Embedding network structure.}%
    \label{fig:embedding_network}%
\end{figure}

First, the preprocessed image is concatenated with the preprocessed watermark. The concatenation is then used as input for the four-block embedding network, as shown in Figure \ref{fig:embedding_network}. The first three blocks include a convolutional layer with 64 filters, a batch normalization, and a ReLU activation function. The last block consists of a convolutional layer with one filter and a Tanh activation function to achieve an appropriate output size.

As mentioned earlier, in the host image preprocessing network, the LL channel was passed to the embedding network. Now we change the main LL with the watermarked one and apply a wavelet inverse transform which results in the watermarked image.

\subsection{Attack Simulator}
\begin{figure}[H]
    \centering
    \includegraphics[width=4.3cm]{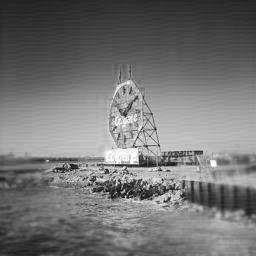}
    \qquad
    \includegraphics[width=4.3cm]{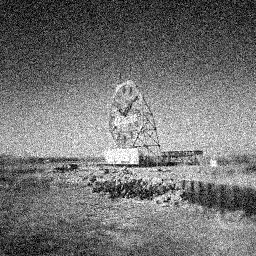}
    \qquad
    \includegraphics[width=4.3cm]{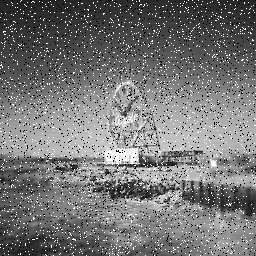} 
    
    \subfloat[\centering Original image]{{\includegraphics[width=4.3cm]{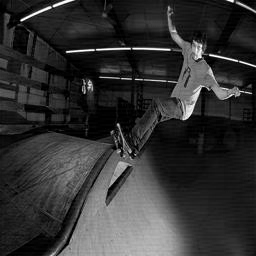} }}%
    \qquad
    \subfloat[\centering Gaussian's attack]{{\includegraphics[width=4.3cm]{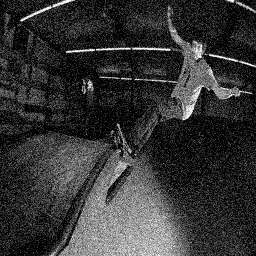} }}%
    \qquad
    \subfloat[\centering Salt \& pepper attack]{{\includegraphics[width=4.3cm]{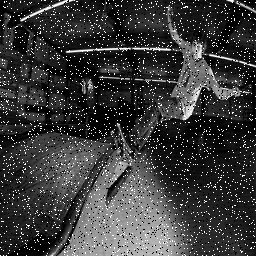} }}%
    \caption{Examples of attacks.}%
    \label{fig:attacks1}%
\end{figure}

Watermarked images might suffer from malicious attacks. In order to increase its robustness, an attack simulator was added to the network. This simulator executes different types of attacks with the ratio in Table \ref{table:ratio} on each mini-batch. Figure \ref{fig:attacks1} shows how explained attacks affect the original image.

\begin{table}[H]
    \centering
    \caption{Attacks for attack simulator.}
    \begin{tabular}{|c|c|c|}
    \hline
    \textbf{Attack} & \textbf{Info} & \textbf{Ratio} \\ \hline
    No attack & - & 1/3 \\ \hline
    Salt and Pepper & $p=0.1$ & 1/6 \\ \hline
    Gaussian noise & $\sigma=0.15$ & 1/6 \\ \hline
    JPEG & quality$=50$ & 1/6 \\ \hline
    Dropout & $p=0.3$ & 1/6 \\ \hline
    \end{tabular}
    \label{table:ratio}
\end{table}

\subsection{Extraction Network}
\begin{figure}[H]
    \centering
    \includegraphics[height=4.5cm]{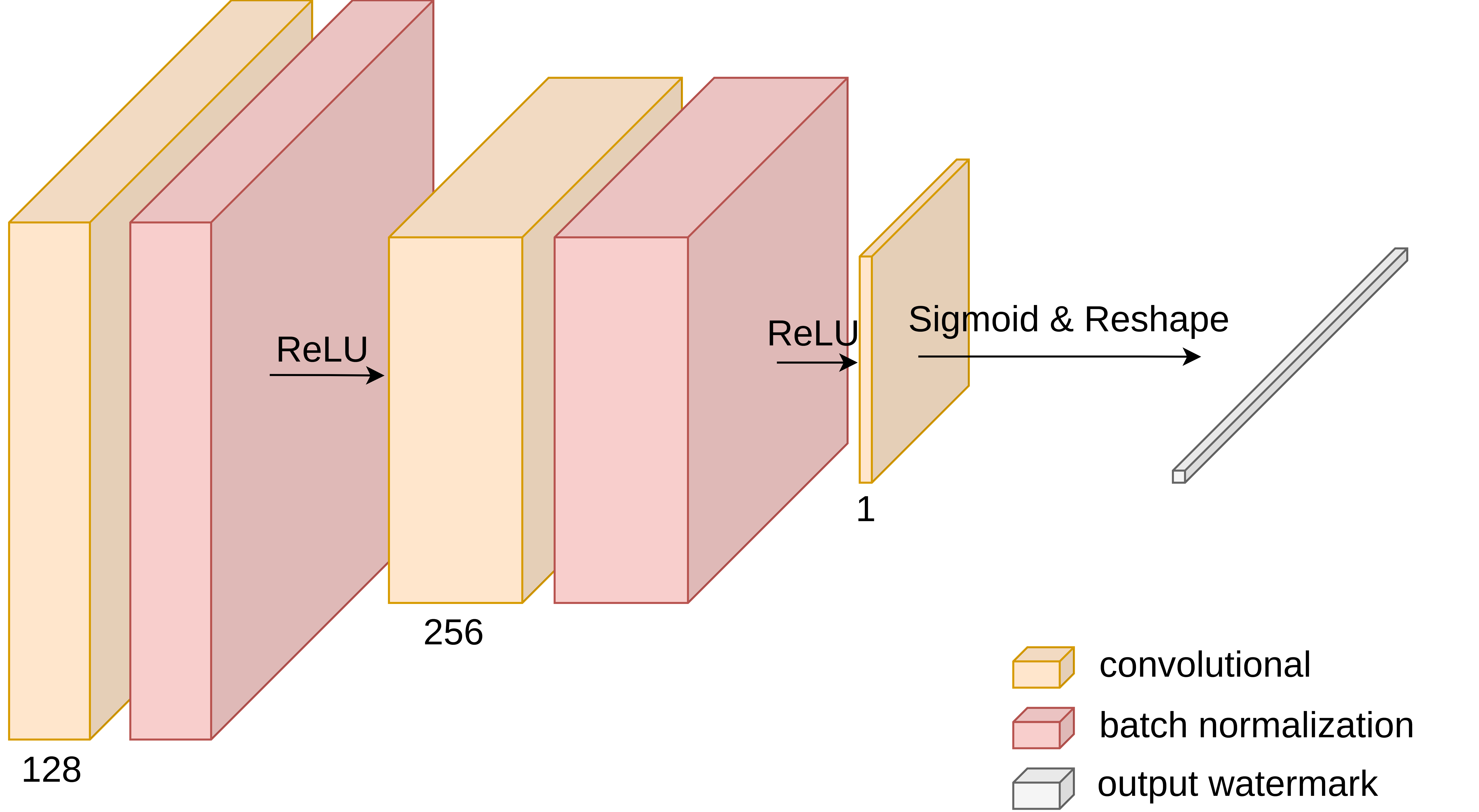}
    \caption{Extraction network structure.}%
    \label{fig:extraction_network}%
\end{figure}

The watermark is embedded in the LL channel, so the first step is to extract the LL channel. The extracted channel is then used as input for the three-block extracting network, as shown in Figure \ref{fig:extraction_network}. The first two blocks include two convolutional layers with 128 and 256 filters, a batch normalization, and a ReLU activation function. The last block consists of a convolutional layer with one filter and a Sigmoid activation function. All convolutional layers in this section will have a stride of 2 to adjust the output shape to the watermark shape.

\subsection{Loss Function}

The two-term loss function is used for training the network. The similarity between a watermarked image and the host image is measured by the mean square error (MSE) loss function shown in Equation \ref{eq:mse}.
\begin{equation}
\label{eq:mse}
    L_1 = \frac{1}{MN} \sum_{i, j}^{MN} [I_{H}(i, j) - I_{WI}(i, j)]^2
\end{equation}
where $M$ and $N$ are the resolution of the host image, $I_{H}$ is the host image, and $I_{WI}$ is the watermarked one.

Also, the mean absolute error (MAE) function is used as the loss function between the extracted watermark and input watermark displayed in Equation \ref{eq:mae}.
\begin{equation}
    \label{eq:mae}
    L_2 = \frac{1}{X} \sum_{i}^{X} |WM_{O}(i) - WM_{E}(i)| 
\end{equation}
where X is the size of the watermark, $WM_{O}$ is the original watermark, and $WM_{E}$ is the extracted one.

As shown in Equation \ref{eq:weightedloss}, a combination of these functions determines the loss function of the whole network.
\begin{equation}
\label{eq:weightedloss}
L_3 = \lambda_1 L_1 +  \lambda_2 L_2
\end{equation}
coefficients ($\lambda_1$ and $\lambda_2$) are hyper-parameters that adjust invisibility and robustness.

\section{Experiments \label{sec:experiments}}
Previously, we described the network in detail. This section explains training materials, such as datasets, training facilities, and evaluation metrics.

\subsection{Dataset}

COCO dataset\footnote{\href{http://images.cocodataset.org/zips/train2017.zip}{http://images.cocodataset.org/zips/train2017.zip}} was used containing 328,000 color images in different sizes \cite{dataset}. Randomly, 40000 images were chosen as training data and 1000 as testing dataset. All images in the dataset were resized to (256, 256, 3) to prepare for feeding into the network. For the watermark, a random generator was used to generate 256-bit watermarks for each instance. By generating random watermarks for each instance, the network became more generalized.

\subsection{Training}
The proposed watermarking network was trained in a Laptop with an Intel(R) Core(TM) i7-9750H CPU @ 2.60 GHz, 32 GB RAM, and the NVIDIA GeForce GTX 1650 GPU. Every mini-batch consists of 10 images and 10 new watermarks. There were 60 epochs. 33.0 for image loss weight  ($\lambda_1$) and 0.2 for watermark loss weight ($\lambda_2$). The learning rate was 0.001.

\subsection{Metrics}
Two metrics were used to estimate robustness and watermark invisibility.
Most previous works used peak-signal-to-noise ratio (PSNR) to evaluate the invisibility of the watermarked image. PSNR measures the similarity between the watermarked image and the host image by Equation \ref{eq:psnr}.
\begin{equation}
    \label{eq:psnr}
    PSNR = 10\log_{10}(\frac{MAX_I^2}{MSE})
\end{equation}
here, $MAX_I$ is the maximum possible pixel value of the image. Also, $MSE$ is the mean square difference between the host image and the watermarked one. 

Due to the importance of the extracted watermark, another metric is required to assess the robustness of the network. In this study, bit error ratio (BER) was used, which is shown in Equation \ref{eq:ber}.
\begin{equation}
    \label{eq:ber}
    \begin{split}
    BER &= \frac{100}{X} \sum_{1 \leq i \leq X} D(WM_{O}(i), WM_{E}(i)) \\
    D(x, y) &= 
    \begin{cases}
    0, & x = y \\
    1, & x \neq y \\
    \end{cases}
    \end{split}
\end{equation}
where $X$ is the size of the watermark, $WM_{O}$ is the original watermark and $WM_{E}$ is the extracted one.

\subsection{Results}
In this section, the algorithm described above is examined from various perspectives. First, the effect of the attack simulator is described. Then there is a comparison between the accuracy of the wavelet channels. In the end, it will be compared to state-of-art methods.

\subsubsection{Attack Simulator Analysis}
In order to evaluate the functionality of the attack simulator, the model was trained without it, and table \ref{table:attackless} shows a significant difference which suggests the critical role of the attack simulator in network robustness.

\begin{table}[H]
\centering
\caption{Effect of the attack simulator on robustness.}
\begin{tabular}{|c|c|c|}
\hline
\textbf{Attack} & \textbf{\begin{tabular}[c]{@{}c@{}}Attack-simulator \\[-1em] free model\end{tabular}} & \textbf{\begin{tabular}[c]{@{}c@{}}Model with \\[-1em] attack simulator\end{tabular}} \\ \hline
PSNR & 50.9 & 40.1 \\ \hline
No attack (BER) & 0.014 & 0.0003 \\ \hline
Salt \& pepper (BER) & 46.44 & 1.4231 \\ \hline
Gaussian noise (BER) & 43.55 & 7.5382 \\ \hline
\end{tabular}
\label{table:attackless}
\end{table}

\subsubsection{Wavelet Channels Analysis}
\begin{figure}[H]
    \centering
    \includegraphics[height=6.5cm]{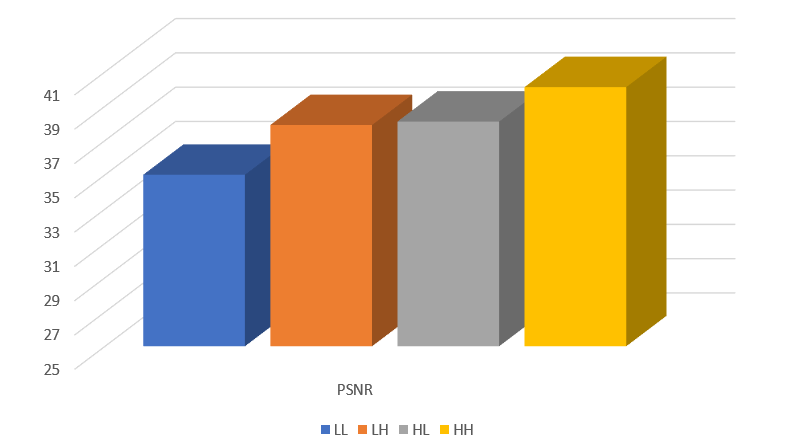}
    \caption{PSNR comparison of wavelet layers.}%
    \label{fig:psnr_comparison}%
\end{figure}

DWT bands have different metrics and efficiency values, as suggested by Shaila R Hallur in \cite{waveletchannel} The purpose of this section is to analyze the results of each band in order to choose the most proficient model. These experiments are based on the same parameters setting with different DWT bands. According to Figure \ref{fig:psnr_comparison}, the HH band is preferred for the PSNR criterion.

\begin{figure}[H]
    \centering
    \includegraphics[height=6.5cm]{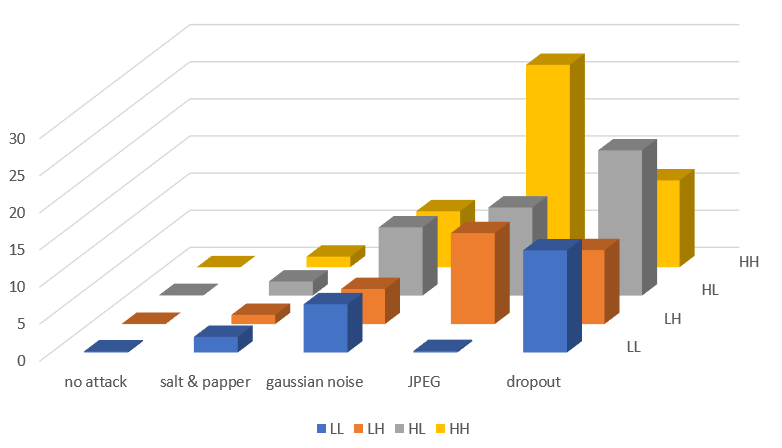}
    \caption{BER comparison of wavelet layers.}%
    \label{fig:ber_comparison}%
\end{figure}

Nevertheless, in the BER comparison, LL has superior performance, as shown in Figure \ref{fig:ber_comparison}.

Further investigations are conducted on the HH band in the remainder of the paper.

\subsubsection{Comparison with State-of-the-Arts Methods}

\begin{figure}[H]
    \centering
    \includegraphics[width=4.3cm]{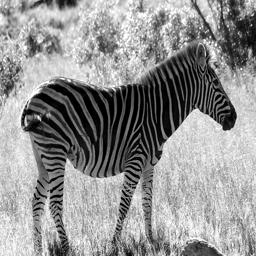}
    \qquad
    \includegraphics[width=4.3cm]{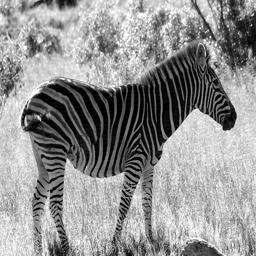}
    \qquad
    \includegraphics[width=4.3cm]{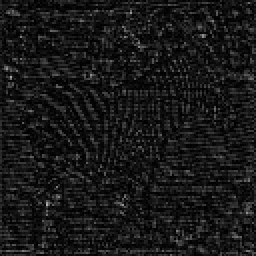} 
    
    \subfloat[\centering host image]{{\includegraphics[width=4.3cm]{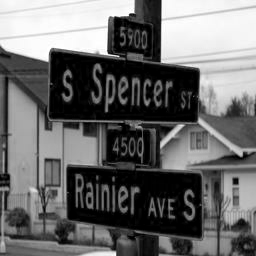} }}%
    \qquad
    \subfloat[\centering watermarked image]{{\includegraphics[width=4.3cm]{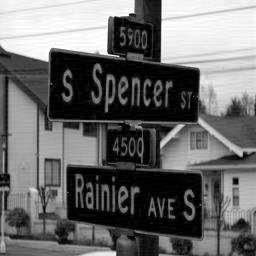} }}%
    \qquad
    \subfloat[\centering 10 times magnified difference of host and watermarked image]{{\includegraphics[width=4.3cm]{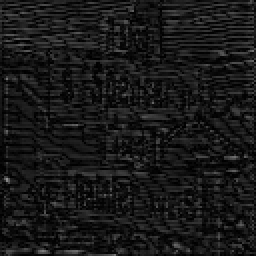} }}%
    \caption{Examples of input images, output images, and their differences.}%
    \label{fig:samples}%
\end{figure}

As mentioned, for evaluating the invisibility of the watermarked images, the PSNR of the watermarked images and the host images on the test dataset are calculated, and the average of these numbers results in 40.1 dB.
As shown in Figure \ref{fig:samples}, the difference between the first and second columns is quite negligible.

The measured BER on output watermarks without any attack is 0.0003. However, applying the attacks, this number increased to 9.58.

Table \ref{table:ber2} shows the network's robustness under each attack. These performances are compared with state-of-art methods. Regarding Table \ref{table:ber2}, using wavelet transform in the structure keeps PSNR, but there is an improvement in the BER metric. Additionally, the proposed network has performed better under Gaussian noise and Salt and Pepper attacks.

\begin{table}[H]
    \centering
    \caption{Comparing proposed and state-of-art methods.}
    \begin{tabular}{|c|c|c|c|}
    \hline
    \textbf{Attack} & \textbf{ReDMark} & \textbf{\cite{seven}} & \textbf{Proposed} \\ \hline
    PSNR & 40.24 & 40.58 & 40.1 \\ \hline
    No attack (BER) & - & 0.7015 & 0.0003 \\ \hline
    Salt and pepper (BER) & 9.1 & 3.1888 & 1.4231 \\ \hline
    Gaussian noise (BER) & 14.5 & 27 & 7.5382  \\ \hline
    Drop out (BER) & 8 & 4.7194 & 11.7051 \\ \hline
    JPEG (BER) & 25.4 & 0.6696 & 27.2482 \\ \hline
    \end{tabular}
    \label{table:ber2}
\end{table}

\section{Conclusion \& Future Work \label{sec:conclusion}}
In this paper, we propose a watermarking neural network that can work well for attacked images and has fewer convolutional layers than previous networks due to wavelet transform. Compared with \cite{seven}, the overall BER has decreased. The PSNR metric of the proposed model is slightly higher than this one; in addition, the model is smaller in size and works better in attacked images.

According to Arpit Bansal in \cite{conclusion}, considering some special attacks in the network leads to poor performance under new attacks. We claimed deterministic methods are inflexible, and thus we used a neural network. However, we are still using deterministic methods in the attack simulator. As a result, the attack simulator can be updated to a more general form in future efforts. It is also possible to investigate famous neural network architectures in order to improve results.

\bibliographystyle{unsrt} 
\bibliography{ref} 
\end{document}